\documentclass[journal,10pt]{IEEEtran}
\usepackage{amssymb}
\usepackage{amsmath}
\usepackage{amsthm}
\usepackage{amsfonts}
\usepackage{accents}
\usepackage{multirow}
\usepackage{array}
\usepackage{multirow}

\usepackage{graphicx}
\usepackage{epstopdf}
\usepackage{float}
\usepackage{pgfplots}
\pgfplotsset{compat=1.8}
\usepackage{epstopdf,epsfig}
\usepackage{subfig}

\usepackage[switch]{lineno}
\usepackage[named]{algo}
\usepackage[noend]{algpseudocode}
\usepackage{algorithm}

\newtheoremstyle{colon}%
{}
{}
{\rm}
{}
{\itshape}
{:}
{ }
{\thmname{#1}\thmnumber{ \itshape#2}\thmnote{ (#3)}}
\theoremstyle{colon}

\begin{document}

\title{Robust Precoding for Rate-Splitting-Based Cell-Free MU-MIMO Networks \vspace{-0.1em}}

\author{André R. Flores and Rodrigo C. de Lamare \vspace{-0.25em}

\thanks{The authors are with the Centre for Telecommunications Studies, Department of Electrical Engineering, Pontifical Catholic University of Rio de Janeiro, Brazil. R. C. de Lamare is with the School of Physics, Engineering and Technology, University of York, United Kingdom. Emails: andre\_flores@esp.puc-rio.br, delamare@puc-rio.br}}

\maketitle
\begin{abstract}
Cell-free (CF) multiuser multiple-input multiple-output (MU-MIMO) systems are an emerging technology that provides service simultaneously to multiple users but suffers from multiuser interference (MUI). In this work, we propose a robust transmit scheme based on rate-splitting (RS) for CF MU-MIMO systems in the presence of imperfect channel state information (CSI) and MUI. We also develop a robust linear precoder design for both private and common precoders based on the minimum mean square error (MMSE) criterion, which incorporates in its design statistical information about the imperfect CSI to provide extra robustness to RS-CF MU-MIMO systems. A statistical analysis is carried out to derive closed-form sum-rate expressions along with a study of the computational complexity of the proposed scheme. Simulation results show that the proposed scheme outperforms conventional robust and non-robust schemes. \vspace{-0.25em}
\end{abstract}

\section{Introduction}
Wireless communications systems are constantly evolving to satisfy the increasing demand for higher data rates, lower latency, and new applications. However, the wireless architecture based on base stations (BSs) with large antenna arrays \cite{mmimo,wence} is unlikely to support the requirements of future applications. Therefore, cell-free (CF) has arisen as a promising technology that is compatible with emerging applications \cite{Elhoushy2021, Ammar2022,Ngo2024}. CF employs multiple access points (APs) distributed over a region of interest to provide service to several user equipments (UEs). Specifically, CF has shown improvements in terms of sum-rate and energy efficiency over conventional networks \cite{Ngo2017,mashdour2022enhanced,mashdour2024clustering,llrref,llridd,oclidd,rspa}. To achieve this gain, multiple antennas or APs are distributed to ensure better channel conditions. Moreover, multiple APs allow the use of well-known signal processing techniques such as precoding and interference cancellation to improve performance. 

Despite its benefits, major problems are found in CF systems: signaling, imperfect CSI, computational cost and MUI. Due to the multiple APs, several channel gains need to be estimated, increasing dramatically the amount of signaling required. Moreover, increasing the number of APs turns into high costs since the complexity scales with the number of APs, which is not suitable for practical systems \cite{Bjoernson2020a}. To overcome both problems, algorithms based on clusters of APs and UEs have been developed \cite{Buzzi2017,Buzzi2020}, which avoid the use of network-wide (NW) techniques that employ all the APs for simultaneous transmission. 
To this end, the system must estimate all the channel gains corresponding to the links between the APs and the users. In contrast, employing clusters of APs reduces the signaling since a small set of channel estimates is required to be conveyed to the APs. Furthermore, a cluster of UEs lowers the computational cost as the calculation of a reduced-dimension precoding matrix is simpler. 

Cluster-based approaches deal with the cost and scalability problems in CF systems, but the MUI caused by the non-orthogonality of received signals and that is further increased by imperfect channel state information (CSI) remains. MUI degrades heavily the performance of CF systems. Therefore, there is great potential for robust techniques capable of effectively operating under imperfect CSI conditions. In \cite{rmmse} a NW robust precoder has been introduced to operate in CF systems. A robust transmission scheme known as rate-splitting (RS) \cite{Mao2022,Clerckx2023,rsrbd,rsthp,rapa,Flores2023} has also been considered with CF systems. In the RS transmission scheme, the messages are split, modulated, and encoded into a common stream and a private stream to further enhance the benefits of the robust approach \cite{Clerckx2023,Mao2022}. Then, power allocation and precoding are performed to transmit the signal. In particular, in \cite{Mishra2022} RS has been incorporated to NW CF massive MIMO for machine-type communications. In \cite{Flores2023}, a cluster-based CF system with RS is proposed where disjoint clusters are considered.

In this work, we present a robust RS-CF MIMO system framework to deal with MUI and increase its robustness against imperfect CSI \cite{rrsprec}. Unlike works that rely only on the RS transmit scheme to increase the robustness of the system, our approach incorporates the design of robust precoders to reap the benefits of RS schemes. Specifically, we devise a robust linear precoder design for both private and common precoders, which incorporates in its design statistical information about imperfect CSI to provide extra robustness to RS-CF MIMO systems. A statistical analysis is carried out to derive closed-form sum-rate expressions along with a study of the computational cost of the proposed scheme. Numerical results show that the proposed scheme outperforms competing techniques. 

The rest of this manuscript is organized as follows. Section II introduces the system model. Section III presents the proposed RS-based robust transmission scheme and the robust precoders. Section IV details the computational cost and the statistical analysis. Section V illustrates the numerical results and Section VI gives the conclusions.

\section{System Model}

Let us consider the downlink of a user-centric CF system with $N_t$ APs distributed over the region of interest. The APs provide service to $K$ users, which are also distributed over the same region of interest. In this work, we consider an underloaded scenario, i.e, $N_t \ge K$. The overall performance of the system drops in overloaded scenarios, but user scheduling can be implemented to mitigate this effect at the expense of extra complexity as detailed in \cite{Mao2021}. Both the users and the APs are equipped with a single omnidirectional antenna. Let us define the channel between the $n$th AP and the $k$th user by
\begin{equation}  g_{n,k}=\sqrt{\zeta_{n,k}}h_{n,k},
\end{equation}
where $\zeta_{n,k}$ and $h_{n,k}$ denote the large-scale fading and small-scale fading coefficient, respectively. In particular, $\zeta_{n,k}$ models the path-loss and shadowing effect and $h_{n,k}$ follows a complex Gaussian distribution with zero mean and unit variance. Since the time division duplex (TDD) protocol is employed the channel can be estimated by pilot training. Then, the estimate of each channel coefficient is described by
\begin{equation}
\hat{g}_{n,k}=\sqrt{\zeta_{n,k}}\left(\sqrt{1+\sigma_e^2}h_{n,k}-\sigma_e\tilde{h}_{n,k}\right), \label{channel_est}
\end{equation}
where $\tilde{h}_{n,k}$ represents the error in the estimate and $\sigma_e \
\in \left[0,1\right]$ can be interpreted as the quality of the channel estimate where, $\sigma_e=0$ represents perfect CSI. It follows that:
\begin{equation}
    g_{n,k}=\frac{1}{\tau}\left(\hat{g}_{n,k}+\tilde{g}_{n,k}\right),
\end{equation}
where $\tau= \sqrt{1+\sigma_e^2}$ and $\tilde{g}_{n,k}=\sigma_e\sqrt{\zeta_{n,k}}\tilde{h}_{n,k}$. The vector $\mathbf{g}_k=\left[g_{1,k},g_{2,k},\cdots,g_{N_t,k}\right]^{\text{T}}\in \mathbb{C}^{N_t}$ is the channel between the APs and the $k$th UE. Then, the channel coefficients can be gathered into the matrix $\mathbf{G}^H=\left[\mathbf{g}_1,\mathbf{g}_2,\cdots,\mathbf{g}_{K}\right]^H\in \mathbb{C}^{K \times N_t}$. Similarly, we have $\hat{\mathbf{G}}^H=\left[\hat{\mathbf{g}}_1,\hat{\mathbf{g}}_2,\cdots,\hat{\mathbf{g}}_{K}\right]^H\in \mathbb{C}^{K \times N_t}$ and $\tilde{\mathbf{G}}^H=\left[\tilde{\mathbf{g}}_1,\tilde{\mathbf{g}}_2,\cdots,\tilde{\mathbf{g}}_{K}\right]^H\in \mathbb{C}^{K \times N_t}$. Thus, we have 
\begin{equation}
        \mathbf{G}^{H}=\frac{1}{\tau}\left(\hat{\mathbf{G}}^{H}+\tilde{\mathbf{G}}^{H}\right).\label{Imperfect Channel model}
\end{equation}
The distributed APs transmit the vector $\mathbf{x}\in\mathbb{C}^{N_t}$, which contains the information intended for all UEs and has a transmit power constraint given by $\mathbb{E}\left[\lVert\mathbf{x}\rVert^2\right]\leq P_t$, where $P_t$ is the transmit power. The received signal is given by
\begin{equation}
\mathbf{y}=\mathbf{G}^{H}\mathbf{x}+\mathbf{n}=\frac{1}{\tau}\hat{\mathbf{G}}^{H}\mathbf{x}+\frac{1}{\tau}\tilde{\mathbf{G}}^{H}\mathbf{x}+\mathbf{n}.\label{Received signal general equation}
\end{equation}
where $\mathbf{n}\in\mathbb{C}^{K}$ represents the additive white Gaussian noise (AWGN) vector with zero mean and variance equal to $\sigma_n^2$. 

\subsection{AP clustering}

Since a small cluster of APs contributes with the most relevant part of the received signal, AP selection is performed, which translates into a reduction in the amount of signaling. Specifically, all the APs serving the $i$th UE are gathered in set $\mathcal{A}_i$. With the sets $\left\{\mathcal{A}_i\right\}_{i=1}^{K}$ defined, we can introduce the sparse channel matrix $\bar{\mathbf{G}}^H=\left[\bar{\mathbf{g}}_1,\bar{\mathbf{g}}_2,\cdots,\bar{\mathbf{g}}_{K}\right]^H\in \mathbb{C}^{K \times N_t}$, where each channel coefficient is defined as follows:
\begin{equation}
    \bar{g}_{n,k}=\begin{cases}
        \hat{g}_{n,k} &n\in \mathcal{A}_k,\\
        0 &\text{Otherwise.}
    \end{cases}
\end{equation}

\subsection{AP selection}
We employ the AP selection adopted in \cite{Flores2022}, which allows us to reduce the signaling load, while keeping the computational complexity low. The APs are selected based on the large-scale fading coefficient. For this purpose, let us define $\mu_\zeta=\frac{1}{K N_t}\sum_{n=1}^{N_t}\sum_{k=1}^{K}\zeta_{n,k}$. Then, we can use the threshold:
\begin{equation}
    \hat{\zeta}_{n,k}=\zeta_{n,k}-\mu_\zeta.
\end{equation}
If $\hat{\zeta}_{n,k}>0$, then $n$ belongs to $\mathcal{A}_k$.

\section{Proposed RS-based Robust Scheme}

Let us consider an RS transmit scheme \cite{Clerckx2023,rsrbd}, where for simplicity the message of user $k$ is split into a common message $m_c$ and a private message $m_{k,p}$. A general RS scheme may split the messages of several users and the common message obtained by splitting the message of a single user is a special case. However, since we aim to evaluate the total ergodic sum-rate splitting the message of a single user is sufficient as explained in \cite{Hao2015}. The common and private messages are encoded, resulting in a common symbol $s_c$ that is superimposed to a vector of private symbols $\mathbf{s}_p=\left[s_1,s_2,\cdots,s_K\right]^{\text{T}}\in \mathbb{C}^{K}$, where $s_k$ contains the private information of the $k$-th user. Thus, the vector of transmit symbols is denoted by $\mathbf{s}^{\left(\text{RS}\right)}=\left[s_c,\mathbf{s}^{\text{T}}_p\right]^{\text{T}}\in \mathbb{C}^{1+K}$.

\begin{figure}[H]
\begin{center}
\includegraphics[width=0.75\columnwidth]{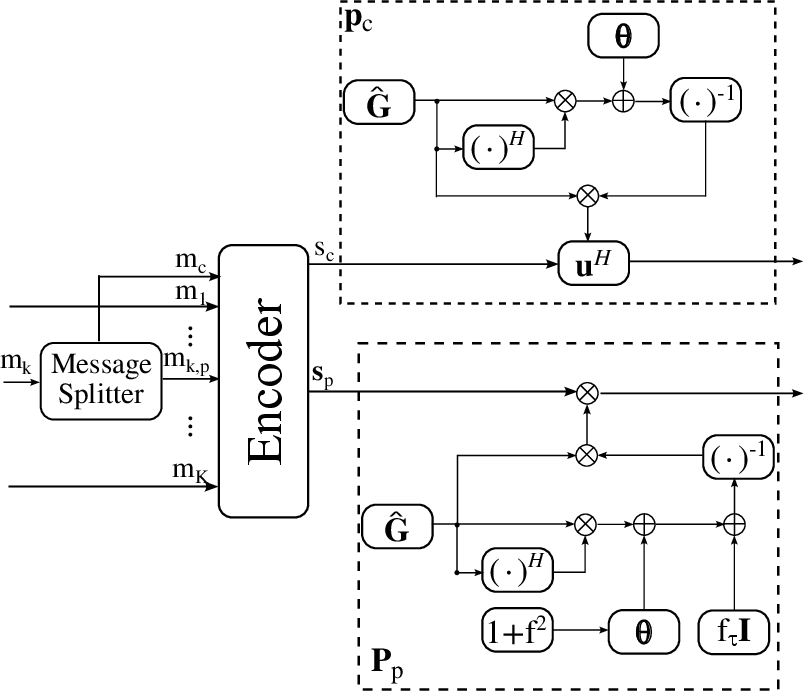}
\caption{Block Diagram of the Proposed Robust Precoder}
\label{Fig:TX}
\end{center}
\end{figure}

At the transmit side, a precoding matrix $\mathbf{P}^{\left(\text{RS}\right)}=\left[\mathbf{p}_{c},\mathbf{P}_p\right]$ is employed to deal with the MUI and enhance the transmission by properly mapping the symbols to the transmit antennas, as depicted in Fig. \ref{Fig:TX}. In particular, $\mathbf{p}_{c}$ maps the common symbol while $\mathbf{P}_{p}$ maps the private symbols. Specifically, each column of $\mathbf{P}_{p}$ maps a private stream to the APs, i.e., $\mathbf{P}_p=\left[\mathbf{p}_1,\mathbf{p}_2,\cdots,\mathbf{p}_K\right]\in \mathbb{C}^{N_t \times K}$ where $\mathbf{p}_k$ maps the $k$-th private symbol to the APs. The transmit vector is given by
\begin{align}
\mathbf{x}&=\mathbf{P}^{\left(\text{RS}\right)}\mathbf{s}^{\left(\text{RS}\right)}=s_c\mathbf{p}_c+\sum_{i=1}^{K}s_i\mathbf{p}_i.
\end{align}
The received vector can be described by
\begin{align}
\mathbf{y}=s_{c}\mathbf{G}^{H}\mathbf{p}_{c}+\sum_{i=1}^{K} s_i\mathbf{G}^H\mathbf{p}_i+\mathbf{n}. \label{rec_vec}
\end{align}
Employing the channel model given by \eqref{Imperfect Channel model} in \eqref{rec_vec}, we obtain
\begin{align}
    \mathbf{y}=&\frac{s_c}{\tau}\hat{\mathbf{G}}^{H}\mathbf{p}_c+\frac{s_c}{\tau}\tilde{\mathbf{G}}^H\mathbf{p}_c+\frac{1}{\tau}\sum_{i=1}^K s_i\hat{\mathbf{G}}^H\mathbf{p}_i\nonumber\\
    &+\frac{1}{\tau}\sum_{j=1}^K s_j\tilde{\mathbf{G}}^H\mathbf{p}_j+\mathbf{n}.
\end{align}
The terms associated with $\tilde{\mathbf{G}}$ are the residual MUI caused by imperfect CSI that should be mitigated by a robust precoder. Here, we consider linear precoders \cite{siprec,Joham2005,gbd,wlbd,robprec,lrcc} but it is possible to extend the work to nonlinear precoders \cite{mbthp,rmbthp,rsthp,bbprec}.

\subsection{Common precoder design}

Let us consider the vector of common symbols $\mathbf{s}_c=\left[s_c,\cdots,s_c\right]^{\text{T}}\in\mathbb{C}^{K}$. Since all users must decode the common stream, then the vector $\mathbf{s}_c$ represents the desired vector of symbols. The proposed robust common precoder can be obtained by solving the following optimization problem: 
\begin{equation}
    \mathbf{p}_c= \arg \min_{\mathbf{p}_c'}J\left(\mathbf{p}_c'\right),\label{Opt problem common precoder}
\end{equation}
where
\begin{equation}
J\left(\mathbf{p}_c'\right)=\underbrace{\mathbb{E}\left[\lVert\mathbf{s}_c-\frac{s_c}{\tau}\hat{\mathbf{G}}^{H}\mathbf{p}_c'\rVert^2\right]}_{T_1}+\mathbb{E}\left[\lVert\underbrace{ \frac{s_c}{\tau}\tilde{\mathbf{G}}^H\mathbf{p}_c'}_{\Delta_c}\rVert^2\right].\label{objective function for the precoder}
\end{equation}
The precoder in \eqref{Opt problem common precoder}, minimizes the difference between the desired vector of common symbols and the transmitted common symbols. Moreover, the proposed common precoder minimizes the residual MUI $\Delta_c$ produced by the imperfect CSI at the transmit side (CSIT). The common symbol is decoded by treating the private information as noise and thus the private streams are not considered in \eqref{Opt problem common precoder}. The residual MUI of the private streams is mitigated by the robust private precoder.

Let us compute the expected value in $T_1$, which yields
\begin{equation}
    T_1= K-\frac{1}{\tau}\mathbf{u}^{\text{T}}\hat{\mathbf{G}}^H\mathbf{p}_c'-\frac{1}{\tau}\mathbf{p}_c'^H\hat{\mathbf{G}}\mathbf{u}+\frac{1}{\tau^2}\mathbf{p}_c'^H\hat{\mathbf{G}}\hat{\mathbf{G}}^{H}\mathbf{p}_c', \label{T1 common precoder}
\end{equation}
where $\mathbf{u}$ is a column vector of length $K$, which contains ones in all of its entries. For the residual MUI, we have
\begin{equation}
    \mathbb{E}\left[\lVert\Delta_c\rVert^2\right]=\frac{1}{\tau^2}\mathbf{p}_c'^H\boldsymbol{\theta}\mathbf{p}_c', \label{T2 common precoder}
\end{equation}
where $\boldsymbol{\theta}=\mathbb{E}\left[\tilde{\mathbf{G}}\tilde{\mathbf{G}}^H\right].$

By substituting \eqref{T1 common precoder} and \eqref{T2 common precoder} in \eqref{objective function for the precoder} and then taking the partial derivative with respect to $\mathbf{p}_c$, we obtain 
\begin{equation}
    \frac{\partial J}{\partial\mathbf{p}_c'^H}=\frac{1}{\tau^2} \hat{\mathbf{G}}\hat{\mathbf{G}}^H\mathbf{p}_c'+\frac{1}{\tau^2}\boldsymbol{\theta}\mathbf{p}_c'-\frac{1}{\tau}\hat{\mathbf{G}}\mathbf{u}.\label{partial derivative common precoder}
\end{equation}
By equating \eqref{partial derivative common precoder} to a null vector and solving for $\mathbf{p}_c$ we obtain the proposed robust common precoder given by
\begin{equation}    \mathbf{p}_c=\tau\left(\hat{\mathbf{G}}\hat{\mathbf{G}}^{H}+\boldsymbol{\theta}\right)^{-1}\hat{\mathbf{G}}\mathbf{u}.
\end{equation}
The robust precoder can be expressed as $\mathbf{p}_c=\alpha_c\bar{\mathbf{p}}_c$, where $\bar{\mathbf{p}}_c$ is a vector with unit norm, i.e., $\bar{\mathbf{p}}_c=\frac{\mathbf{p}_c}{\lVert\mathbf{p}_c\rVert}$. It follows that $\alpha_c$ is the power allocated to the common stream. Note that the robust precoder can also be extended to multiple common streams, which requires one robust precoder per stream.

\subsection{Private Precoder design}

Once the common symbols are decoded, successive interference cancellation (SIC) removes its contribution from the received signal. The received signal after SIC is given by 
\begin{equation}
    \mathbf{y}=\frac{1}{\tau}\hat{\mathbf{G}}^{H}\mathbf{P}_p\mathbf{s}_p+\underbrace{\frac{1}{\tau}\tilde{\mathbf{G}}^{H}\mathbf{P}_p\mathbf{s}_p}_{\boldsymbol{\Delta_p}}+\mathbf{n}.
\end{equation}

The optimal private precoder minimizes the effect of the residual MUI (e.g., by letting $\mathbb{E}\left[\lVert\boldsymbol{\Delta}\rVert^2\right]\to 0$). Such a precoder can be obtained by solving the following optimization: \vspace{-0.25em}
\begin{equation}
\begin{split}
    \left\{\mathbf{P}_p,f\right\}=& \text{argmin}~~\underbrace{\mathbb{E}\left[\lVert\mathbf{s}_p-f^{-1}\mathbf{y}\rVert^2\right]+\mathbb{E}\left[\lVert\boldsymbol{\Delta}_p\rVert^2\right]}_{J_p}\nonumber\\    
    & \hspace{-1em} \text{subject to}~~\mathbb{E}\left[\lVert\mathbf{x}\rVert^2\right]=\text{tr}\left(\mathbf{P}_p\mathbf{P}_p^H\right)=P_t-\alpha_c,\label{minimization problem MMSE-RB}
\end{split}
\end{equation}
\vspace{-0.25em}
where $\text{tr}(\cdot)$ is the trace of its matrix argument and $f$ can be interpreted as an automatic gain control at the receivers \cite{Joham2005,rmmse}. Expanding the terms of $J_p$ and evaluating $J_p$, we get
\begin{align}
    J_p&=\nonumber\\& K-\frac{f^{-1}}{\tau}\text{tr}\left(\mathbf{P}_p^H\hat{\mathbf{G}}\right)-\frac{f^{-1}}{\tau}\text{tr}\left(\hat{\mathbf{G}}^H\mathbf{P}_p\right)+\frac{1}{\tau^2}\text{tr}\left(\boldsymbol{\theta}\mathbf{P}_p\mathbf{P}_p^H\right)\nonumber\\
    &+\frac{f^{-2}}{\tau^2}\left[\text{tr}\left(\mathbf{P}_p\mathbf{P}_p^H\hat{\mathbf{G}}\hat{\mathbf{G}}^H\right)+\text{tr}\left(\mathbf{P}_p\mathbf{P}_p^H\boldsymbol{\theta}\right)\right]+f^{-2}\text{tr}\left(\mathbf{R}_n\right).
\end{align}
Then, the Lagrangian function of the optimization problem is 
\begin{align}
    \mathcal{L}(\mathbf{P}_p,f,&\lambda)=\nonumber\\&K-\frac{f^{-1}}{\tau}\left[\text{tr}\left(\mathbf{P}_p^H\hat{\mathbf{G}}\right)+\text{tr}\left(\hat{\mathbf{G}}^{H}\mathbf{P}_p\right)\right]+f^{-2}\text{tr}\left(\mathbf{R}_\mathbf{n}\right)\nonumber\\
    &+\frac{f^{-2}}{\tau^2}\left[\text{tr}\left(\mathbf{P}_p\mathbf{P}_p^H\hat{\mathbf{G}}\hat{\mathbf{G}}^H\right)+\text{tr}\left(\mathbf{P}_p\mathbf{P}_p^H\boldsymbol{\theta}\right)\right]\nonumber \\ 
    &+\frac{1}{\tau^2}\text{tr}\left(\boldsymbol{\theta}\mathbf{P}_p\mathbf{P}_p^{H}\right)+\lambda\left[\text{tr}\left(\mathbf{P}_p\mathbf{P}_p^H\right)-P_t+\alpha_c\right].\label{Lagrangian of the robust MMSE}
\end{align}
Computing the partial derivatives of the Lagrangian and equating them to zero, we obtain
\vspace{-0.5em} 
\begin{equation}    
\mathbf{P}_p=f\tau{\bar{\mathbf{P}}},
\end{equation}
    \begin{equation}
        \lambda=\frac{K\sigma_n^2}{f^2\left(P_t-\alpha_c\right)}-\frac{\textup{tr}\left(\mathbf{P}_p^{H}\boldsymbol{\theta}\mathbf{P}_p\right)}{\tau^2\left(P_t-\alpha_c\right)},\label{solution for lambda}
    \end{equation}
\vspace{-0.5em}
where
\vspace{-0.5em}
\begin{equation}
{\bar{\mathbf{P}}}=(\underbrace{\hat{\mathbf{G}}\hat{\mathbf{G}}^H+\left(1+f^2\right)\boldsymbol{\theta}+\lambda f^2\tau^2\mathbf{I}}_{\mathbf{P}^{\left(i\right)}})^{-1}\hat{\mathbf{G}},\label{p_bar robust mmse}
\end{equation}
    \begin{equation}
    f=\frac{1}{\tau}\sqrt{\frac{P_t-\alpha_c}{\text{tr}\left(\bar{\mathbf{P}}\bar{\mathbf{P}}^{H}\right)}},\label{power scaling factor}
\end{equation}
assuming that the inverse of $\mathbf{P}^{\left(i\right)}$ exists.

Note that $\mathbf{P}_p$ depends on $\lambda$ and vice-versa. To obtain them, we employ an alternating optimization framework, where one of the variables is fixed while the other is computed. We begin the computations with the standard MMSE precoder as the initial state. With the initial precoder, we update the parameter $\lambda$ iteratively. The design procedure is given in Algorithm \ref{alg:Robust MMSE}. 
\begin{algorithm}[H]
	\caption{Proposed MMSE Robust Precoders}
	\label{alg:Robust MMSE}
    \begin{algorithmic}[1]
    \Statex \textbf{Input:}  $\hat{\mathbf{G}}^{H}$, $\mathbf{u}$, $P_t$, $\sigma_n^2$ ,$\sigma_e^2$, $\alpha_c$, $i_t$
    \Statex \textbf{Output:} $\mathbf{P}^{\left(\text{RS}\right)}$
\Statex Robust Common Precoder
\State $\tau \leftarrow \sqrt{1+\sigma_e^2}$ 
\State
$\boldsymbol{\Theta} \leftarrow \mathbb{E}\left[\tilde{\mathbf{G}}\tilde{\mathbf{G}}^H\right]$
\State $\mathbf{p}_c\leftarrow\tau\left(\hat{\mathbf{G}}\hat{\mathbf{G}}^{H}+\boldsymbol{\theta}\right)^{-1}\hat{\mathbf{G}}\mathbf{u}$

\Statex Robust Private Precoder
\State $\bar{\mathbf{P}}\left[0\right]\leftarrow\left(\hat{\mathbf{G}}\hat{\mathbf{G}}^{H}+\frac{K\sigma_n^2}{P_t-\alpha_c}\mathbf{I}\right)^{-1}\hat{\mathbf{G}}$
\State
$f\left[0\right]\leftarrow\frac{1}{\tau}\sqrt{\frac{P_t-\alpha_c}{\text{tr}\left(\bar{\mathbf{P}}\left[0\right]\bar{\mathbf{P}}^H\left[0\right]\right)}}$
\State
$\mathbf{P}_p\left[0\right]\leftarrow f\left[0\right]\tau\bar{\mathbf{P}}\left[0\right]$
\State
$\lambda\left[0\right]\leftarrow\frac{K\sigma_n^2}{f^2\left[0\right] (P_t-\alpha_c)}-\frac{\text{tr}\left(\mathbf{P}^{H}_p\left[0\right]\boldsymbol{\theta}\mathbf{P}_p[0]\right)}{\tau^2\left(P_t-\alpha_c\right)}$
\For{$i=1:i_t$}
\State
$\bar{\mathbf{P}}\left[i\right] \leftarrow(\hat{\mathbf{G}}\hat{\mathbf{G}}^{H}+\left(1+f^2\left[i-1\right]\right)\boldsymbol{\theta}+\lambda\left[i-1\right] f^2\left[i-1\right]\tau^2\mathbf{I} )^{-1}\hat{\mathbf{G}}$
\State
$f\left[i\right]\leftarrow\frac{1}{\tau}\sqrt{\frac{P_t-\alpha_c}{\text{tr}\left(\bar{\mathbf{P}}\left[i\right]\bar{\mathbf{P}}^H\left[i\right]\right)}}$
\State 
$\mathbf{P}_p\left[i\right]\leftarrow f\left[i\right]\tau\bar{\mathbf{P}}\left[i\right]$
\State $\lambda\left[i\right]\leftarrow \frac{K\sigma_n^2}{f^2\left[i\right] (P_t-\alpha_c)}-\frac{\text{tr}\left(\mathbf{P}^{H}_p\left[i\right]\boldsymbol{\theta}\mathbf{P}_p[i]\right)}{\tau^2\left(P_t-\alpha_c\right)}$
\EndFor
\State \textbf{end for}
\State
$\mathbf{P}_p\leftarrow f\left[i_t\right]\tau\bar{\mathbf{P}}\left[i_t\right]$
\State $\mathbf{P}^{\left(\text{RS}\right)}\leftarrow\left[\mathbf{p}_c,\mathbf{P}_p\right]$
\Statex \Return $\mathbf{P}^{\left(\text{RS}\right)}$
\end{algorithmic}\label{alg1 robust MMSE}
\end{algorithm}

\subsection{{Power allocation}}
{An efficient way to allocate power to the common  and private streams is to compute the sum-rate in the saturation region of the conventional precoder without RS. The power is allocated to the private precoder such that the private sum-rate equals the performance of the conventional precoder without RS, while the remaining power is allocated to $\alpha_c$ \cite{Clerckx2023}.}

\section{Statistical Analysis and Computational Cost}

In this section, we present a stastistical analysis that derives closed-form expressions for the ergodic sum-rate (ESR) and the computational cost of the proposed robust precoder.

\subsection{Ergodic Sum-Rate Performance}

The ESR of the proposed scheme is given by
\begin{equation}
     S_e=\mathbb{E}\left[\bar{R}_{c,k}\right]+\sum_{l=1}^K \mathbb{E}\left[\bar{R}_l\right].\label{system ergodic sum rate}
 \end{equation}
 where $\bar{R}_{c,k}=\mathbb{E} \left[R_{c,k} \left(\mathbf{G}^H\right)|\mathbf{\hat{G}}^H\right]$ denotes the average common rate and  $\bar{R}_{k}=\mathbb{E}\left[R_{k}\left(\mathbf{G}^H\right)|\mathbf{\hat{G}}^H\right]$ stands for the average private rate of user $k$, and assuming Gaussian signalling $R_{c,k}=\log_2\left(1+\gamma_{c,k}\right)$ and $R_k=\log_2\left(1+\gamma_k\right)$ denote the instantaneous common rate and the instantaneous private rate, respectively. The signal-to-interference-plus-noise ratio (SINR) when decoding the common stream at the $k$-th user is 

\begin{align}
\gamma_{c,k}&=\frac{\alpha_c^2\lvert \hat{\mathbf{g}}_k^{H}\boldsymbol{\Psi}\mathbf{u}\rvert^2}{\alpha_c^2\delta_{c,k}+\tau^2\lVert\boldsymbol{\Psi}\mathbf{u}\rVert^2\left(f^2\sum\limits_{i=1}^K \lvert \boldsymbol{g}_k^{H}\bar{\mathbf{p}}_i\rvert^2+\sigma_n^2\right)},
\end{align}
where $\delta_{c,k}=2\Re\left\{{\left(\mathbf{u}^{\text{T}}\boldsymbol{\Psi}^H\hat{\mathbf{g}}_k\right)\left(\tilde{\mathbf{g}}_k^H\boldsymbol{\Psi}\mathbf{u}\right)}\right\}+\lvert\tilde{\mathbf{g}}^H\boldsymbol{\Psi}\mathbf{u}\rvert^2$, $\boldsymbol{\Psi}=\left(\hat{\mathbf{G}}\hat{\mathbf{G}}^{H}+\boldsymbol{\theta}\right)^{-1}\hat{\mathbf{G}}$, and $\Re\left\lbrace \cdot \right\rbrace$ denotes the real part of a complex argument. On the other hand, the SINR when decoding the private stream of the $k$-th user is
\begin{align}
\gamma_k&=\frac{f^2\lvert\hat{\mathbf{g}}_k^{\normalfont{H}}\bar{\mathbf{p}}_k\rvert^2}{f^2\sum\limits_{\substack{i=1\\i\neq k}}^K \lvert\hat{\mathbf{g}}_k^{H}\bar{\mathbf{p}}_i\rvert^2+f^2\sum\limits_{j=1}^K\delta_{j,k}+\sigma_n^2},\label{instantaneous SINR private rate perfect}
 \end{align}
 where $\delta_{j,k}=2\Re\left\{{\left(\bar{\mathbf{p}}_j^H\hat{\mathbf{g}}_k\right)\left(\tilde{\mathbf{g}}_k^H\bar{\mathbf{p}}_j\right)}\right\}+\lvert\tilde{\mathbf{g}}_{k}^H\bar{\mathbf{p}}_j\rvert^2$.
 
\subsection{Computational Cost}

We employ the total number of floating points operations (FLOPs) to describe the computational cost as follows:
\begin{itemize}
    \item Multiplication of $l\times m$ and $m \times n$ complex matrices requires $8lmn-2ln$ FLOPs.
    \item The inverse of a complex square matrix of size $m \times m$ requires $\frac{4}{3}m^3$ FLOPs.
    \item To obtain $\mathbf{p}_c$, a total of $\frac{4}{3}N_t^3+8N_t^2K+8N_tK-4N_t+1$ FLOPs are required.
    \item In the first iteration, $\bar{\mathbf{P}}\left[0\right]\leftarrow\left(\hat{\mathbf{G}}\hat{\mathbf{G}}^{H}+\frac{K\sigma_n^2}{P_t-\alpha_c}\mathbf{I}\right)^{-1}\hat{\mathbf{G}}$ requires a total of $\frac{4}{3}N_t^3+16N_t^2K-2N_t^2-2N_tK+2N_t+3$ FLOPS, $f\left[0\right]\leftarrow\frac{1}{\tau}\sqrt{\frac{P_t-\alpha_c}{\text{tr}\left(\bar{\mathbf{P}}\left[0\right]\bar{\mathbf{P}}^H\left[0\right]\right)}}$ requires $8N_tK+2$ FLOPs, $\mathbf{P}_p\left[0\right]\leftarrow f\left[0\right]\tau\bar{\mathbf{P}}\left[0\right]$ requires $N_tK+1$ FLOPs and $\lambda\left[0\right]\leftarrow\frac{K\sigma_n^2}{f^2\left[0\right] (P_t-\alpha_c)}-\frac{\text{tr}\left(\mathbf{P}^{H}_p\left[0\right]\boldsymbol{\theta}\mathbf{P}_p[0]\right)}{\tau^2\left(P_t-\alpha_c\right)}$ requires $8N_t^2K+6N_tK+6$.
    
    \item For any other iteration $i\neq 0$, the computational cost of $f\left[i\right]$, $\mathbf{P}_p\left[i\right]$  $\lambda\left[i\right]$ remains the same as the cost of $f\left[0\right]$, $\mathbf{P}_p\left[0\right]$  $\lambda\left[0\right]$, respectively. In contrast, $\bar{\mathbf{P}}\left[i\right] \leftarrow(\hat{\mathbf{G}}\hat{\mathbf{G}}^{H}+\left(1+f^2\left[i-1\right]\right)\boldsymbol{\theta}+\lambda\left[i-1\right] f^2\tau^2\left[i-1\right]\mathbf{I} )^{-1}\hat{\mathbf{G}}$ requires $\frac{4}{3}N_t^3+16N_t^2K+2N_t^2-2N_tK+2N_t+3$ FLOPs.
\end{itemize}
The computational cost in terms of FLOPS is given by
\begin{align}
    C_f&=
    i_t\left(\frac{4}{3}N_t^3+24N_t^2K+2N_t^2+13N_tK+2N_t+12\right)\nonumber\\&
    +\frac{8}{3}N_t^3+32N_t^2K+21N_tK-2N_t^2-2N_t+12,
\end{align}
where $i_t$ denotes the number of iterations. The conventional MMSE and RS-MMSE precoders have a computational complexity of $\mathcal{O}\left(N_t^3\right)$, but both approaches obtain a worse performance under imperfect CSI than the proposed approach. 
\section{Simulations}
 Numerical examples were performed to evaluate the performance of the proposed robust techniques. To this end, the large scale fading coefficients are given by $\zeta_{k,n}=P_{k,n}\cdot 10^{\frac{\sigma^{\left(\textrm{s}\right)}z_{k,n}}{10}}$,
 where $P_{k,n}$ denotes the path loss, and the shadowing effect is described by $z_{k,n}$ with standard deviation  equal to $\sigma^{\left(\textrm{s}\right)}=8$. $z_{k,n}$ follows a Gaussian distribution with zero mean and unit variance. The path loss was calculated using a three-slope model \cite{Palhares2020} described by \par\noindent\small
 \begin{align}
     P_{k,n}=\begin{cases}
  -L-35\log_{10}\left(d_{k,n}\right), & \text{$d_{k,n}>d_1$} \\
  -L-15\log_{10}\left(d_1\right)-20\log_{10}\left(d_{k,n}\right), & \text{$d_0< d_{k,n}\leq d_1$}\\
    -L-15\log_{10}\left(d_1\right)-20\log_{10}\left(d_0\right), & \text{otherwise,}
\end{cases}
 \end{align}\normalsize
 where $d_{k,n}$ is the distance between the $n$-th AP and the $k$-th user, $d_1=50$ m, $d_0= 10$ m, and the attenuation $L$ is \par\noindent
 \begin{align}
 L=&46.3+33.9\log_{10}\left(f\right)-13.82\log_{10}\left(h_{\textrm{AP}}\right)\nonumber\\
     &-\left(1.1\log_{10}\left(f\right)-0.7\right)h_u+\left(1.56\log_{10}\left(f\right)-0.8\right),
 \end{align}\normalsize
 where $h_{\textrm{AP}}=15$ m and $h_{u}=1.65$ m are the positions of the APs and UEs above the ground, respectively. We consider a frequency of $f= 1900$ MHz. The noise variance is $\sigma_n^2=T_o k_B B N_f,$
 where $T_o=290$ K is the noise temperature, $k_B=1.381\times 10^{-23}$ J/K is the Boltzmann constant, $B=50$ MHz is the bandwidth and $N_f=10$ dB is the noise figure. The signal-to-noise ratio (SNR) is 
$\text{SNR}=\frac{P_{t}\textrm{Tr}\left(\mathbf{G}^{\text{H}}\mathbf{G}\right)}{N_t K \sigma_n^2}.$
Among the CF techniques examined here are the proposed RSCF scheme with a robust private precoder and a standard common precoder denoted by RSCF-MMSE-RB+PpRB, the proposed RSCF scheme with a robust common precoder and a robust private precoder called RSCF-MMSE-RB+PcRB, the RSCF scheme with a standard linear MMSE precoder denoted by RSCF-MMSE, the conventional CF MU-MIMO scheme with a robust MMSE precoder \cite{Palhares2020}, the conventional CF MU-MIMO scheme with a standard linear MMSE precoder, {and the conventional weighted MMSE (WMMSE) precoder \cite{Christensen2008}.}

We consider a small cluster of a CF system in which 12 APs provide service to 3 users. A total of 10000 trials were performed to compute the ESR. In particular, for each channel estimate, we consider 100 different error matrices to compute the ASR. The parameter $\alpha_c$ was computed by employing an exhaustive search with a grid of step size equal to 0.005.
 
\begin{figure}
\centering
\subfloat[]{\includegraphics[width=0.475\linewidth]{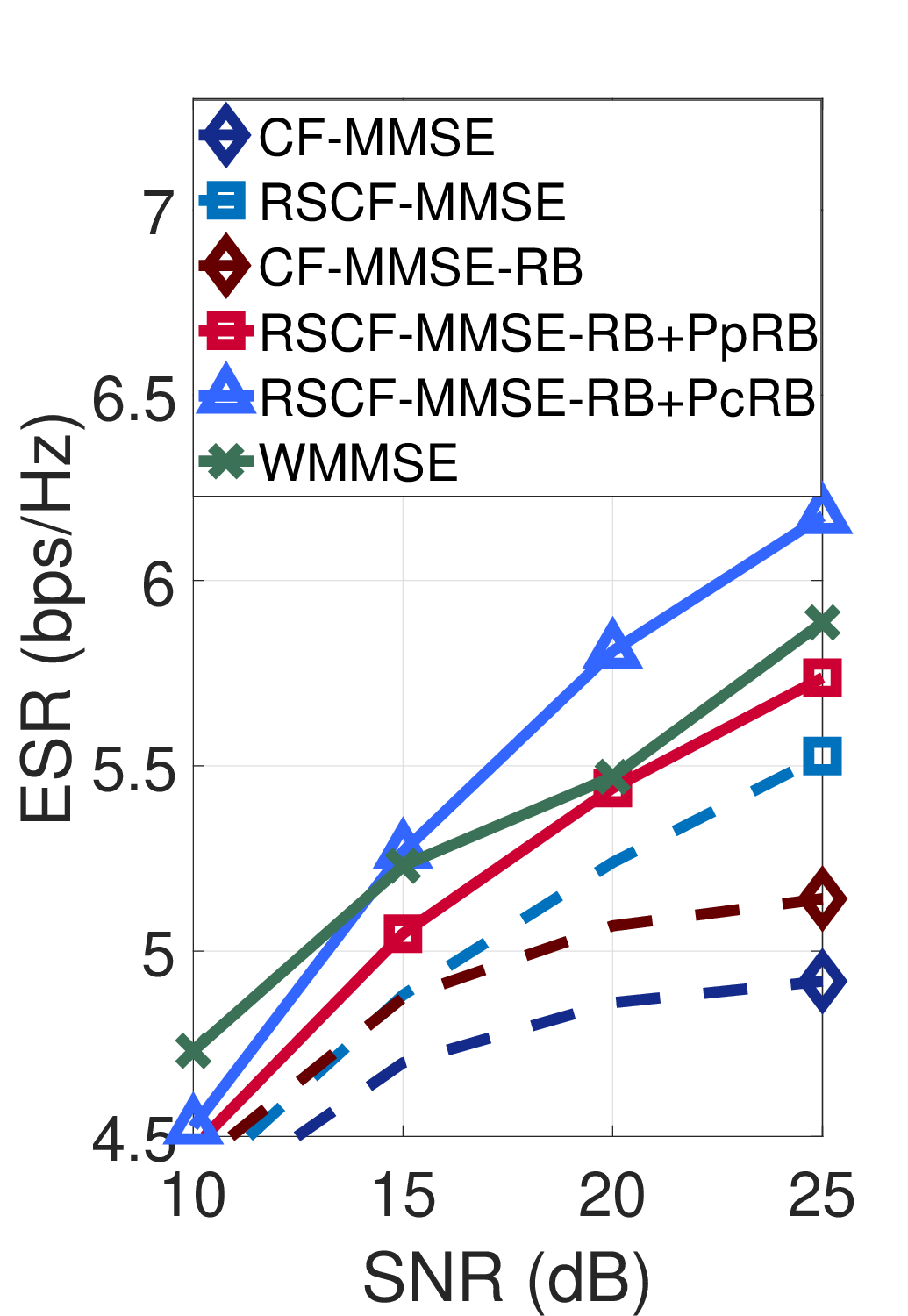}%
\label{ESR vs SNR}}
\hfil
\subfloat[]{\includegraphics[width=0.475\linewidth]{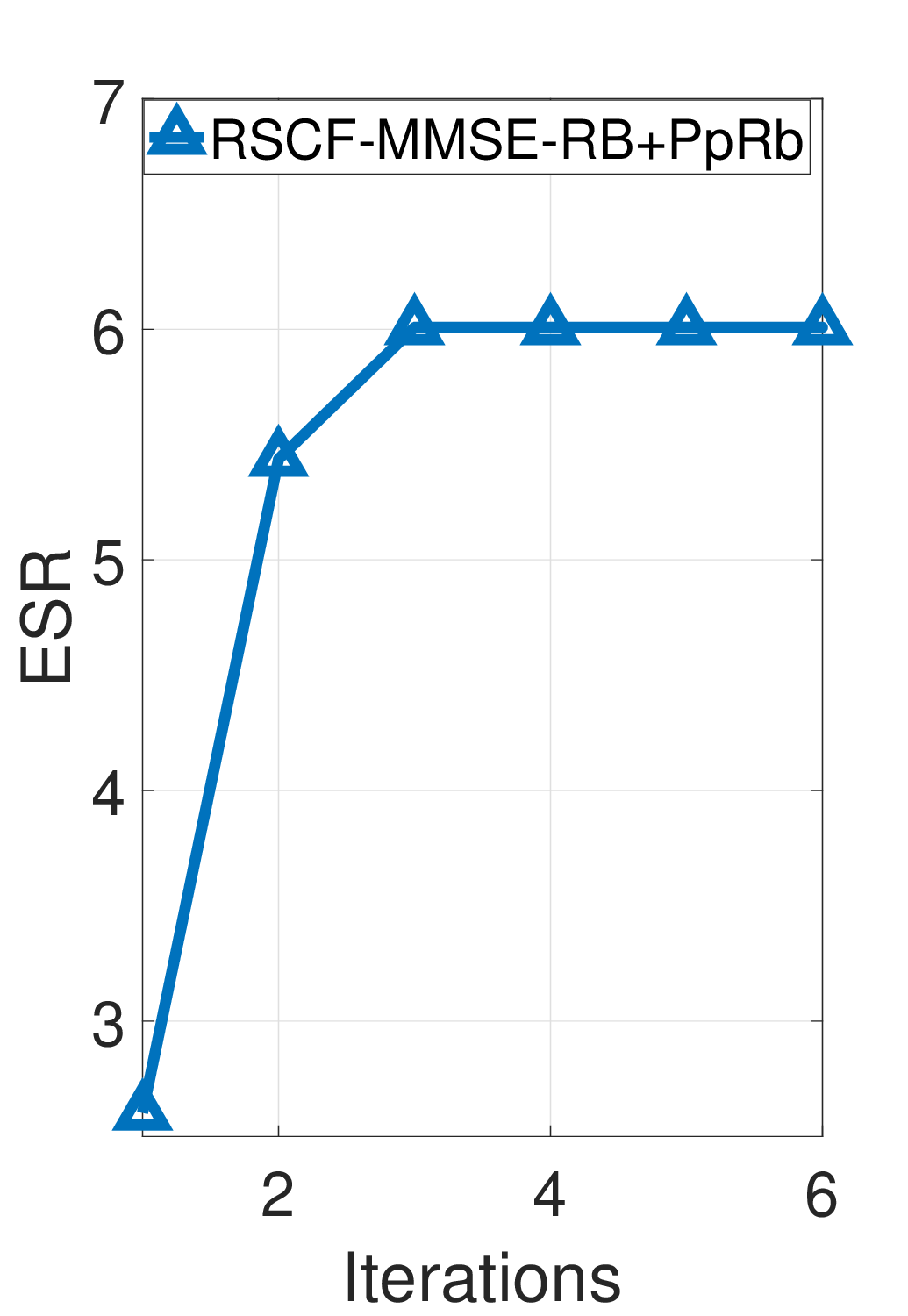}%
\label{ESR vs iterations}}
\caption{Sum-rate performance (a) vs SNR. (b) vs iterations.}
\label{ESR performance}
\end{figure}

 In the first example, we evaluate the performance of the proposed robust schemes in terms of ESR. We set the quality of the channel estimate in \eqref{channel_est} to $\sigma_e^2=0.3$. From Fig. \ref{ESR vs SNR} we can notice that the robust approach performs better than the conventional linear approaches. In particular, the gain increases at high SNR. Indeed, CSIT imperfections degrade heavily the performance  of conventional approaches at high SNR  since the residual MUI scales with the transmit power. The proposed robust RSCF-MMSE-RB+PpRB scheme can effectively deal with CSIT imperfections, providing a substantial gain over other techniques. Furthermore, including a robust common precoder as done by the proposed robust RSCF-MMSE-RB+PcR scheme significantly increases the ESR. Fig. \ref{ESR vs iterations} shows the ESR obtained at $22$ dB by the proposed RSCF-MMSE-RB+PcRB as the number of iterations increases. We consider a random matrix as the initial precoder to test the convergence. The proposed algorithm has a fast convergence, reaching its steady state with only three iterations.

\begin{figure}[H]
\begin{center}
\includegraphics[width=0.95\columnwidth]{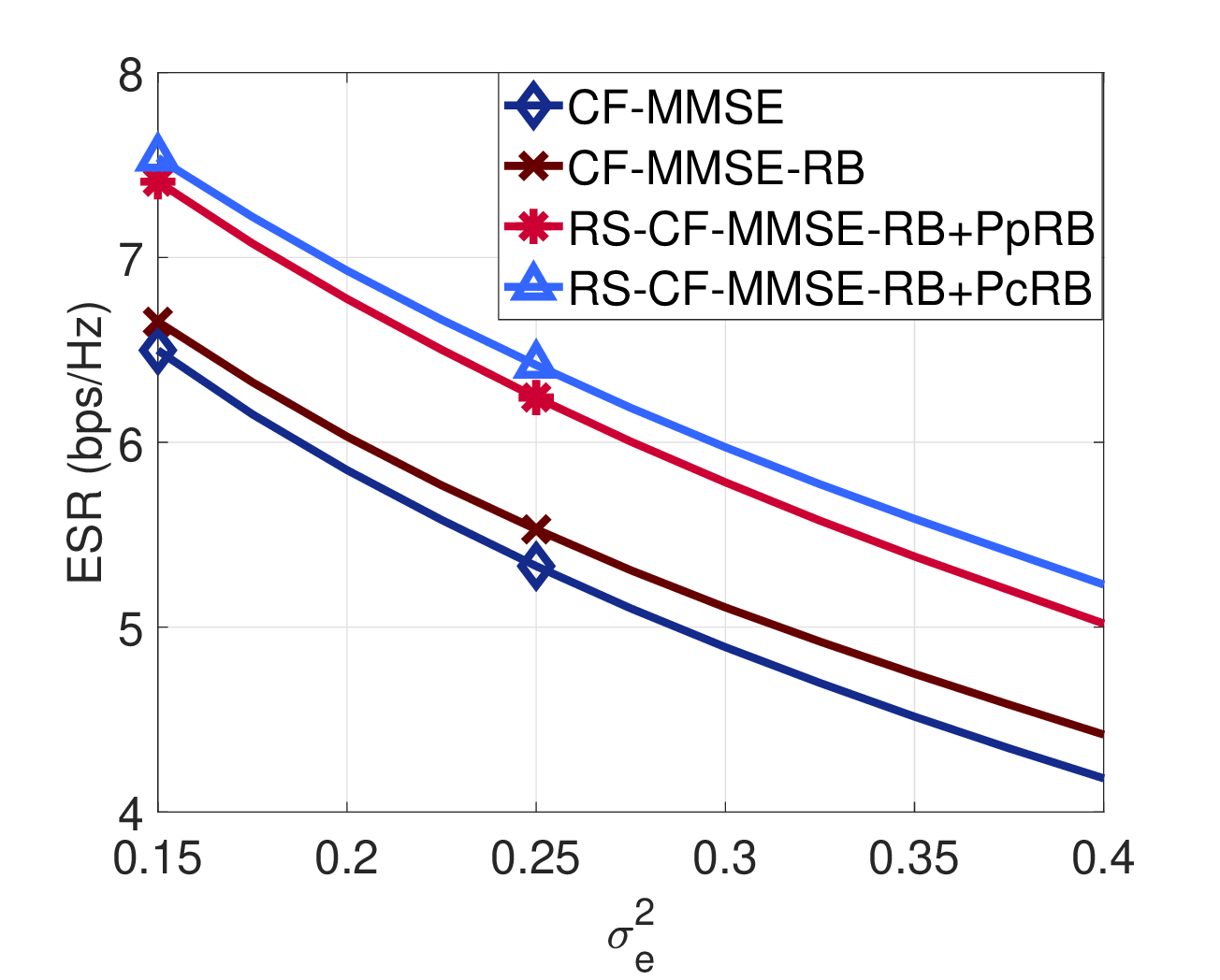}\caption{Performance of Robust MMSE schemes for RS-CF systems with different CSIT quality levels, $N_t=12$, $K=3$.}\label{Fig3}
\vspace{-0.5em}
\end{center}
\end{figure}

In the second example, we show the performance of the proposed robust approaches against different CSIT quality levels in Fig. \ref{Fig3}. For this purpose the SNR was set to $22$ dB. The proposed RSCF-MMSE-RB+PcRB scheme obtains the best performance among the precoders as expected. The robust common precoder is introduced to deal with CSIT imperfections and, therefore, the power allocated increases as the residual MUI gets higher. 

\section{Conclusions}

We have proposed a novel robust transmit scheme, which employs RS along with robust private and common precoders to enhance the performance of CF systems operating under imperfect CSIT. The precoders were derived employing the MMSE criterion and have low computational complexity. Simulation results show that the proposed robust approaches outperform the existing CF-MMSE-RB and RSCF-MMSE schemes. In particular, the inclusion of a robust common precoder brought significant improvements in terms of ESR. \vspace{-0.5em}

\bibliographystyle{IEEEtran}
\bibliography{ref}
\end{document}